\documentclass[showpacs,aps,twocolumn]{revtex4}
\usepackage{amssymb}
\usepackage{amsmath}
\usepackage{graphicx}
\usepackage{color}
\usepackage[colorlinks,hyperindex]{hyperref}

\hypersetup{
colorlinks,
citecolor=blue,
linkcolor=black,
urlcolor=black,
}
\newcommand{\be}{\begin{equation}}
\newcommand{\ee}{\end{equation}}
\newcommand{\ben}{\begin{eqnarray}}
\newcommand{\een}{\end{eqnarray}}
\newcommand{\bes}{\begin{subequations}}
\newcommand{\ees}{\end{subequations}}

\newcommand{\bfi}{\begin{figure}}
\newcommand{\efi}{\end{figure}}
\newcommand{\bc}{\begin{center}}
\newcommand{\ec}{\end{center}}

\begin{document}

\title{Configurational entropy in $f(R,T)$ brane models\medskip }
\author{R. A. C. Correa$^{1,}$\footnote{%
rafael.couceiro@ufabc.edu.br}, and P. H. R. S. Moraes$^{2,}\medskip $\footnote{%
moraes.phrs@gmail.com}}
\affiliation{$^{1}$CCNH, Universidade Federal do ABC, 09210-580, Santo Andr\'{e}, SP,
Brazil\medskip }
\affiliation{$^{2}$ITA - Instituto Tecnol\'{o}gico de Aeron\'{a}utica, 12228-900, S\~{a}o
Jos\'{e} dos Campos, S\~{a}o Paulo, Brazil}

\begin{abstract}
In this work we investigate generalized theories of gravity in the so-called
configurational entropy (CE) context. We show, by means of this
information-theoretical measure, that a stricter bound on the parameter of $%
f(R,T)$ brane models arises from the CE. We find that these bounds are
characterized by a valley region in the CE profile, where the entropy is
minimal. We argue that the CE measure can open a new role and an important
additional approach to select parameters in modified theories of gravitation.
\end{abstract}

\maketitle

\section{Introduction}

Although $\Lambda $CDM cosmological model provides a great match between
theory and observation \cite{hinshaw/2013}, what makes it to be usually
referred to as the \textquotedblleft concordance model", a number of
shortcomings, as the cosmological constant and hierarchy problems, still
await for convincing explanations. While by assuming general relativity as
the gravitational theory, those problems cannot be solve straightforwardly,
higher order derivative and extradimensional theories might contribute
efficiently to solve these issues.

Note that $f(R)$ and $f(R,T)$ cosmological models \cite{sotiriou/2010}-\cite%
{moraes/2015b}, with $R$ and $T$ being respectively the Ricci scalar and the
trace of the energy-momentum tensor, are able to describe the cosmic
acceleration our universe is undergoing (check \cite{riess/1998}-\cite%
{perlmutter/1999}) with no need of invoking a cosmological constant. On the
other hand, the hierarchy problem can be solved by assuming our observable
universe is a sub-manifold which is embedded in an anti-de Sitter
five-dimensional space (AdS$_{5}$), called bulk, as in Randall-Sundrum
braneworld model \cite{RS}.

In this work we will deal with braneworld models in the presence of scalar
fields \cite{GW}-\cite{C2}. Departing from the original Randall and Sundrum
proposal, which leads to a thin braneworld scenario, the coupling with
scalar fields leads to thick braneworlds. The presence of a scalar field
makes the warp function to behave smoothly, yielding such a thickness. This
possibility has opened a new area of study, and here we quote Refs.\cite{G}-%
\cite{G8} for some works on the subject.

Specifically, we will consider the $f(R,T)$ gravity in such a thick
braneworld scenario. Note that although $f(R)$ brane models have already
been presented in the literature \cite{Bazeia-Lobao-Menezes-Petrov-Silva}-%
\cite{liu/2011}, due to its recent elaboration, $f(R,T)$ gravity still lacks
of a significant number of applications in the braneworld. Anyhow, it is
remarkable the fact that in \cite{bazeia/2015}, the authors have pioneered
such an approach.

Here, rather than a cosmological approach, we will investigate $f(R,T)$
brane models from the configurational entropy (CE) perspective. Gleiser and
Stamatopoulos (GS) have proposed in \cite{Gleiser-Stamatopoulos} such a new
physical quantity, which brings additional informations about some
parameters of a given model for which the energy density is localized. They
have shown that the higher the energy that approximates the actual solution,
the higher its relative CE, which is defined as the absolute difference
between the actual function CE and the trial function CE. The CE is able to
solve situations where the energies of the configurations are degenerate. In
this case, it can be used to select the best system configuration.

Although it has been recently proposed, the CE has already been used to
study the non-equilibrium dynamics of spontaneous symmetry breaking \cite%
{PRDgleiser-stamatopoulos}, to obtain the stability bound for compact
objects \cite{Gleiser-Sowinski}, to investigate the emergence of localized
objects during inflationary preheating \cite{PRDgleiser-graham} and to
distinguish configurations with energy-degenerate spatial profiles \cite%
{Rafael-Dutra-Gleiser}. Solitons, Lorentz symmetry breaking, supersymmetry,
and entropy, were employed using the CE concept \cite{Rafael-Roldao,
RafaelPRD, AHEP, RafaelPRD2}. The CE for travelling solitons reveals that
the best value of the parameter responsible for breaking the Lorentz
symmetry is that where the energy density is distributed equally around the
origin. It was argued that the information-theoretical measure of travelling
solitons in Lorentz symmetry violation scenarios can be very important to
probe situations where the parameters responsible for breaking the
symmetries are arbitrary, with the CE selecting the best value of the
parameter in the model. Another interesting work about CE was presented in 
\cite{stamatopoulos/2015}, where it is responsible for identifying the
critical point in the context of continuous phase transitions. Finally, in
braneworld scenarios it was shown that CE can be employed to demonstrate a
high organizational degree in the structure of the system configuration for
large values of a parameter of the sine-Gordon model \cite{bc}. The entropic
information-measure in the context of $f(R)$ braneworlds with non-constant
curvature has also been studied \cite{Rafael-Pedro}.

In this work, we calculate the CE in $f(R,T)$ brane models. We obtain its
profile and reveal the information content of such models. The paper is
organized as follows. In the next section we will review the concept of CE
measure. In Section III, we outline the basic theoretical structure for
generalized $f(R,T)$ theories of gravity. In Section IV, we show two
distinct models in $f(R,T)$ theory and its solutions. In Section V, we
describe the calculations of the CE for $f(R,T)$ theories. We also show the
corresponding CE profile. Finally, in Section VI, we present our conclusions
and directions for future work.

\section{An overview about configurational entropy measure}

Here, we will review the very recent work by GS \cite{Gleiser-Stamatopoulos}%
, where it was shown that, in analogy with Shannon's information entropy, a
CE measure in a functional space can be used to select the best fit solution
with spatially-localized energy. Another interesting consequence highlighted
by GS is the fact that the CE relates the dynamical and informational
content of physical models. In this case, this information-theoretical
measure can be able to extract information about the different solutions
which is related to their spatial profiles. Thus, following \cite%
{Gleiser-Stamatopoulos}, let us begin by writing the CE as

\begin{equation}
\sigma _{c}[f]=-\int d^{d}\mathbf{\omega }\tilde{f}(\mathbf{\omega })\ln [%
\tilde{f}(\mathbf{\omega })],  \label{19}
\end{equation}

\noindent where $d$ is the number of spatial dimensions, $\tilde{f}(\mathbf{%
\omega })=f(\mathbf{\omega })/f_{\max }(\mathbf{\omega })$, where $f_{\max }(%
\mathbf{\omega })$ is the maximal modal fraction, that is, the mode giving
the highest contribution. In this case, the function $f(\mathbf{\omega })$
which was defined as the modal fraction is represented as

\begin{equation}
f(\mathbf{\omega })=\frac{\left\vert \mathcal{F}(\mathbf{\omega }
)\right\vert ^{2}}{\int d^{d}\mathbf{\omega }\left\vert \mathcal{F}(\mathbf{%
\omega })\right\vert ^{2}}.  \label{20}
\end{equation}

The function $\mathcal{F}(\mathbf{\omega })$ represents the Fourier
Transform of the energy density of the configuration. It is important to
remark that the energy density must be square-integrable since in such cases
the entropy can be well defined.

We will extend the procedure presented in \cite{Gleiser-Stamatopoulos},
which is absolutely general when applied to systems with spatially-localized
energy for a scalar field theory that describes generalized gravity coupled
to a scalar field in five-dimensional space-time. Following that work, it is
possible to obtain the entropy of the configurations assuming that $\mathcal{%
F}(\omega )$ obeys the Fourier transform:

\begin{equation}
\mathcal{F}(\omega )=\frac{1}{\sqrt{2\pi }}\int_{-\infty }^{\infty
}dxe^{i\omega y}\rho (y).  \label{1.1.1}
\end{equation}

We are working with spatially-localized energy densities, which are part of
a set of square-integrable bounded functions $\rho (y)\in L^{2}(\mathbf{R})$
and their Fourier transforms $\mathcal{F}(\omega )$. It is also important to
remark that the Plancherel's theorem states that 
\begin{equation}
\int_{-\infty }^{\infty }dy\left\vert \rho (y)\right\vert ^{2}=\int_{-\infty
}^{\infty }d\omega \left\vert \mathcal{F}(\omega )\right\vert ^{2}.
\label{1.1.3}
\end{equation}

As argued by GS, the information-entropic measure approach can be capable of
taking into account the dynamical and the informational contents of models
with localized energy configurations. In this case, the CE provides a
complementary perspective to situations where strictly energy-based
arguments are inconclusive. In fact, as pointed out in \cite%
{Gleiser-Stamatopoulos}, higher CE correlates with higher energy, and
consequently the configuration is more disordered. Therefore, the
information-theoretical measure is responsible to point out which solution
is the most ordered one among a family of infinite solutions.

Thus, we will apply this new approach to investigate generalized gravity
theories coupled to a real scalar field. As we will see, important
consequences will arise from the CE concept. Furthermore, we will show that
the CE provides a stricter bound on the parameters of the $f(R,T)$ models.

\section{$f(R,T)$ theories of gravity}

\label{sec:gen}

In this section, we present a brief review about $f(R,T)$ theories of
gravity. In this context, it was presented some years ago by Harko and
collaborators the seminal work entitled $f(R,T)$ gravity \cite{harko/2011}.
In that context the authors showed that it is possible to construct a
modified theory of gravity, where the gravitational Lagrangian is given by
an arbitrary function of the Ricci scalar $R$ and of the trace of the
energy-momentum tensor $T$. On a physical viewpoint, the dependence on $T$
may be originated by the presence of exotic imperfect fluids or quantum
effects. Therefore, as shown in \cite{harko/2011}, in $(3+1)$ dimensions the
action $S$ for the $f(R,T)$ modified theories of gravity can be written in
the form 
\begin{equation}
S=\frac{1}{16\pi }\int d^{4}x\sqrt{-g}f(R,T)+\int d^{4}x\sqrt{-g}\mathcal{L}
_{m},  \label{adc1}
\end{equation}
where $f(R,T)$ is the arbitrary function of $R$ and $T$, $g$ is the
determinant of the metric $g_{\mu \nu }$ with $\mu $, $\nu $ assuming the
values $0,$ $1,$ $2,$ $3$, and $\mathcal{L}_{m}$ is the matter Lagrangian
density. Note that we are using the natural system of units with $G=c=1$, so
that the Einstein gravitational constant is defined as $\kappa ^{2}=8\pi $.

We emphasize that the $f(R,T)$ modified theory of gravity has been analysed
in the literature in several scenarios. For instance, we can find studies in
cosmological models \cite{moraes/2014}-\cite{moraes/2015b},\cite{ad1}, in
quantum chromo dynamics \cite{ad2}, in a Bianchi Type-II string with
magnetic field \cite{ad3} and in thermodynamics \cite{ad4,harko/2014}.

On the other hand, recently, Bazeia and collaborators, motivated by the
works \cite{harko/2011, ad5}, have found thick brane models in $f(R,T)$
generalized theories of gravity in an $AdS_{5}$ warped geometry with an
extra dimension of infinite extent \cite{bazeia/2015}. It has been shown
that two distinct choices of the functional form of $f(R,T)$ induce
quantitative modifications in the thick brane profile, without changing its
qualitative behavior. Another interesting result obtained in \cite%
{bazeia/2015} is the fact that the gravity sector remains linearly stable in
each choice of $f(R,T)$ analized.

The theory used by Bazeia and collaborators for describing generalized
gravity coupled to a scalar field in five-dimensional space-time, with an
extra dimension $y$ of infinite extent, is given by the following action 
\begin{equation}
S=\int d^{4}x\,dy\sqrt{|g|}\left[ -\frac{1}{4}f(R,T)+\mathcal{L}_{s}\right] ,
\label{eq1}
\end{equation}%
where $\mathcal{L}_{s}$ is the Lagrange density, $g=det(g_{ab})$, with the
indices $a$ and $b$ running from $0$ to $4$, the signature of the metric is $%
(+----)$ and we are using units such that $4\pi G^{(5)}=1$. In this case,
the Lagrangian density is put into the form 
\begin{equation}
\mathcal{L}_{s}=\frac{1}{2}\nabla _{a}\phi \nabla ^{a}\phi -V(\phi )\,.
\label{eq2}
\end{equation}

From the equation above, the energy-momentum tensor of the scalar field in
this theory is given by 
\begin{equation}
T_{ab}=-\frac{1}{2}g_{ab}\nabla _{c}\phi \nabla ^{c}\phi +g_{ab}V+\nabla
_{a}\phi \nabla _{b}\phi \,.  \label{eq3}
\end{equation}

In addition, the corresponding equation of motion for the scalar field and
the modified Einstein equation can be written as 
\begin{eqnarray}
&&\left. \nabla _{a}\nabla ^{a}\phi +\frac{3}{4}\nabla _{a}\left(
f_{T}\nabla ^{a}\phi \right) +\frac{5}{4}V_{\phi }f_{T}+V_{\phi }=0,\right.
\\
&&  \notag \\
&&\left. f_{R}R_{ab}-\frac{1}{2}g_{ab}f+\left( g_{ab}\Box -\nabla _{a}\nabla
_{b}\right) f_{R}\right.  \notag \\
&&  \notag \\
&&\left. =2T_{ab}+\frac{3}{2}f_{T}\nabla _{a}\phi \nabla _{b}\phi \,,\right.
\end{eqnarray}
where $V_{\phi }\equiv dV/d\phi $, $f_{R}\equiv df/dR$ and $f_{T}\equiv
df/dT $. \ 

Following the approach presented in \cite{bazeia/2015}, we will work with
the metric 
\begin{equation}
ds^{2}=e^{2A}\eta _{\mu \nu }dx^{\mu }dx^{\nu }-dy^{2}\,,  \label{eq7}
\end{equation}

\noindent where $y$ is the extra dimension, $\eta _{\mu \nu }$ is the usual
Minkowski metric in the four-dimensional spacetime with signature $(+,-,-,-)$%
, and $e^{2A}$ is the so-called warp factor. In particular, when the warp
function $A$ and the field $\phi $ are static and depend only on the extra
dimension $y$, the equation of motion and the modified Einstein equations
become 
\begin{equation}
\Big(1+\frac{3}{4}f_{T}\Big)\phi ^{\prime \prime }+\Big[(4+3f_{T})A^{\prime
}+\frac{3}{4}f_{T}^{\prime }\Big]\phi ^{\prime }=\Big(1+\frac{5}{4}f_{T}%
\Big) V_{\phi }\,,  \label{eq8}
\end{equation}
\begin{subequations}
\label{eq9}
\begin{eqnarray}
\frac{2}{3}\phi ^{\prime 2}\Big(\!1\!+\!\frac{3}{4}f_{T}\!\Big)\!\!\!
&=&\!\!\!-A^{\prime \prime }f_{R}+\frac{1}{3}A^{\prime }f_{R}^{\prime }- 
\frac{1}{3}f_{R}^{\prime \prime }\,,  \label{eq9a} \\
\!\!\!\!\!\!V(\phi )\!-\!\frac{\phi ^{\prime 2}}{2}\Big(\!1\!+\!\frac{3}{2}
f_{T}\!\Big)\!\!\! &=&\!\!\!2(A^{\prime 2}\!\!+\!A^{\prime \prime
})f_{R}\!-\!\frac{f}{4}\!-\!2A^{\prime }f_{R}^{\prime },  \label{eq9b}
\end{eqnarray}

\noindent where the prime denote derivatives with respect to the extra
dimension.

Here, it is important to highlight that this particular approach, where we
are considering the static case with $A=A(y)$ and $\phi =\phi (y)$, makes
possible the finding of innovative analytical solutions for the theory. A
good example are the interesting solutions found in \cite{bazeia/2015}. In
that work, two distinct models for $f(R,T)$ theories of gravity have been
studied, and it was shown that the gravity sector of the thick braneworld
configurations remains linearly stable. Moreover, each model induces new
quantitative modifications in the thick brane profile. Thus, in order to
make the present work more comprehensive and didactical, in the next section
we will present those two examples and their solutions.

\section{Two specific models and their solutions}

In the present section, we consider two particular classes of $f(R,T)$
modified gravity models, which were investigated in \cite{bazeia/2015}. In
that reference, it has been found that in such particular cases, both the
field $\phi $ and warp function $A$ can be obtained analytically.

\subsection{Case 1: $f(R,T)=R-\protect\alpha \,T^{\,n}$}

In the first case, the $f(R,T)$ function is given by 
\end{subequations}
\begin{equation}
f(R,T)=R-\alpha \,T^{\,n},  \label{eq23}
\end{equation}
where $\alpha $ and $n$ are real parameters. In this case, when $n=1$ the
exact solutions presented in \cite{bazeia/2015} for the field $\phi $ and
the warp function $A$ are written as

\begin{eqnarray}
\phi (y) &=&\frac{1}{b}\arcsin \Big[\tanh (By)\Big]\,,  \label{eq33} \\
A(y) &=&\frac{2\gamma a}{3B}~\ln \Big[\mbox{sech}(By)\Big]\,.  \label{eq34}
\end{eqnarray}%
\noindent where $\gamma $, $a$ and $b$ are real parameters \cite{bazeia/2015}%
, and $B\equiv 4\gamma ab^{2}/(4-3\alpha )$. Furthermore, the energy density
can be put in the form 
\begin{eqnarray}
\rho (y)\!\!\! &=&\!\!\!\frac{16(\gamma a)^{2}}{3(4\!-\!5\alpha )}\!\Big\{%
\!\!-\!1\!+\!\!\Big[\!1\!+\!\frac{3B(1\!-\!\alpha )}{\gamma a(4\!-\!3\alpha )%
}\!\Big]\mbox{sech}^{2}(By)\!\Big\}\!\!\times   \notag \\
\!\!\! &&\!\!\!\Big[\mbox{sech}(By)\Big]^{\frac{4\gamma a}{3B}}\,.
\label{den1}
\end{eqnarray}

\subsection{Case 2: $f(R,T)=R+\protect\beta R^{2}-\protect\alpha T$}

Now, let us investigate another model, where $f(R,T)$ is represented by 
\begin{equation}
f(R,T)=R+\beta R^{2}-\alpha T\,,  \label{eq42}
\end{equation}
with $\alpha $ and $\beta $ being real parameters. In this case the
solutions for the field and warp function can be written in the following
form 
\begin{eqnarray}
\phi (y) &=&\pm \sqrt{\frac{6}{4\!-\!3\alpha }\!-\!\frac{320\beta k^{2}}{%
4-3\alpha }}\mathcal{E}\left( \Phi ,\Xi \right) \,, \\
A(y) &=&\ln \Big[\mbox{sech}(ky)\Big]\,,
\end{eqnarray}
where $k$ is a positive parameter and $\mathcal{E}(\Phi ,\Xi )$ is the
elliptic integral of the second kind with 
\begin{equation}
\Phi \equiv \phi _{s}=\arcsin \Big[\tanh (ky)\Big]\text{, and }\Xi \equiv 
\frac{392\beta k^{2}}{160\beta k^{2}\!-\!3}\text{.}
\end{equation}
Here, the energy density is given by 
\begin{eqnarray}
\rho (y)\!\! &=&\!\!-\frac{12k^{2}}{4\!-\!5\alpha }\tilde{S}^{2}\!+\!\frac{%
12k^{2}(6\!-\!5\alpha )}{(4\!-\!3\alpha )(4\!-\!5\alpha )}\tilde{S}^{4}\!-\!%
\frac{8\beta k^{4}}{4\!-\!5\alpha }\tilde{S}^{2}\Big[10-  \notag \\
\!\! &&\!-\frac{116(6-5\alpha )}{4\!-\!3\alpha }\tilde{S}^{2}\!+\!\frac{%
98(8-7\alpha )}{4\!-\!3\alpha }\tilde{S}^{4}\Big]\,.
\end{eqnarray}

\noindent where $\tilde{S}=\mbox{sech}(ky)$.

\section{CE in $f(R,T)$ theories}

In this section, we will study the entropic profile of the generalized
gravity theories here presented.

\subsection{CE for $f(R,T)=R-\protect\alpha \,T^{\,n}$}

For this first case, we can rewrite the energy density (\ref{den1}) in the
following form 
\begin{equation}
\rho (y)=\sum\limits_{j=1}^{2}A_{j}\mbox{sech}^{\delta _{j}}(By),
\label{eq.1}
\end{equation}

\noindent where $\delta _{1}\equiv \delta $ and $\delta _{2}\equiv \delta +2$%
, with $\delta \equiv 4\gamma a/B$. Furthermore, we have 
\begin{eqnarray}
A_{1} &\equiv &-\frac{16(\gamma a)^{2}}{3(4-5\alpha )},  \label{eq.2} \\
A_{2} &\equiv &-\frac{16(\gamma a)^{2}}{3(4-5\alpha )}\left[ 1+\frac{%
3B(1-\alpha )}{\gamma a(4-3\alpha )}\right] .  \label{eq.3}
\end{eqnarray}
Therefore substituting the energy density given by Eq.(\ref{eq.1}) into Eq.(%
\ref{1.1.1}), we can obtain, after some arduous calculation, the following
Fourier transform 
\begin{equation}
\mathcal{F}(\omega )=\sum\limits_{j=1}^{2}\sum\limits_{\ell =1}^{2}A_{j,\ell
}\text{ }_{2}\mathcal{H}_{1}[\delta _{j},\mu _{j,\ell };\mu _{j,\ell }+1;-1],
\label{eq.4}
\end{equation}

\noindent where $_{2}\mathcal{H}_{1}[\delta _{j},\mu _{j,\ell };\mu _{j,\ell
}+1;-1]$ is the hypergeometric function, and we are using the definitions 
\begin{eqnarray}
\mu _{j,\ell } &\equiv &[B\delta _{j}-(-1)^{\ell +1}i\omega ]/2B,
\label{eq.5} \\
A_{j,\ell } &\equiv &\frac{2^{\delta _{j}-1/2}A_{j}[B\delta _{j}+(-1)^{\ell
+1}i\omega ]}{\sqrt{\pi }(B^{2}\delta _{j}^{2}+\omega ^{2})}.  \label{eq.6}
\end{eqnarray}
Moreover, we have

\begin{equation}
\int_{-\infty }^{\infty }d\omega \left\vert \mathcal{F}(\omega )\right\vert
^{2}=\frac{1}{B}\sum\limits_{j=1}^{2}\sum\limits_{\ell =1}^{2}\frac{%
2^{\delta _{j,\ell }-1}\Im _{j,\ell }\Gamma ^{2}(\delta _{j,\ell }/2)}{%
\Gamma (\delta _{j,\ell })},  \label{eq.7}
\end{equation}

\noindent with 
\begin{equation}
\delta _{j,\ell }\equiv \delta _{j}+\delta _{\ell },\text{ \ }\Im _{j,\ell
}\equiv A_{j}A_{\ell }.
\end{equation}

Now, we can write the modal fraction in the form

\begin{equation}
f(\omega )=B\left[ \frac{\sum\limits_{j,\ell ,m,n=1}^{2}A_{j,\ell
}A_{m,n}^{\ast }\text{ }_{2}\mathcal{H}_{1}^{(j,\ell )}\text{ }_{2}\mathcal{H%
}_{1}^{(m,n)\ast }}{\sum\limits_{j,\ell =1}^{2}\frac{2^{\delta _{j,\ell
}-1}\Im _{j,\ell }\Gamma ^{2}(\delta _{j,\ell }/2)}{\Gamma (\delta _{j,\ell
})}}\right] ,  \label{eq.8}
\end{equation}

\noindent where 
\begin{eqnarray}
_{2}\mathcal{H}_{1}^{(j,\ell )} &=&_{2}\mathcal{H}_{1}[\delta _{j},\mu
_{j,\ell };\mu _{j,\ell }+1;-1], \\
_{2}\mathcal{H}_{1}^{(m,n)} &=&_{2}\mathcal{H}_{1}[\delta _{m},\mu
_{m,n};\mu _{m,n}+1;-1].
\end{eqnarray}

Figure 1 below depicts $f(\omega )$ for different values of $\alpha $. As it
can be seen, the modal fraction profile is influenced by the $\alpha $
parameters. In this case, we can check that the $\alpha $ parameter is
responsible for the appearance of a split in the modal fraction, where there
are two peaks which are symmetric with respect to the origin.

\begin{figure}[h]
\begin{center}
\includegraphics[width=8.7cm]{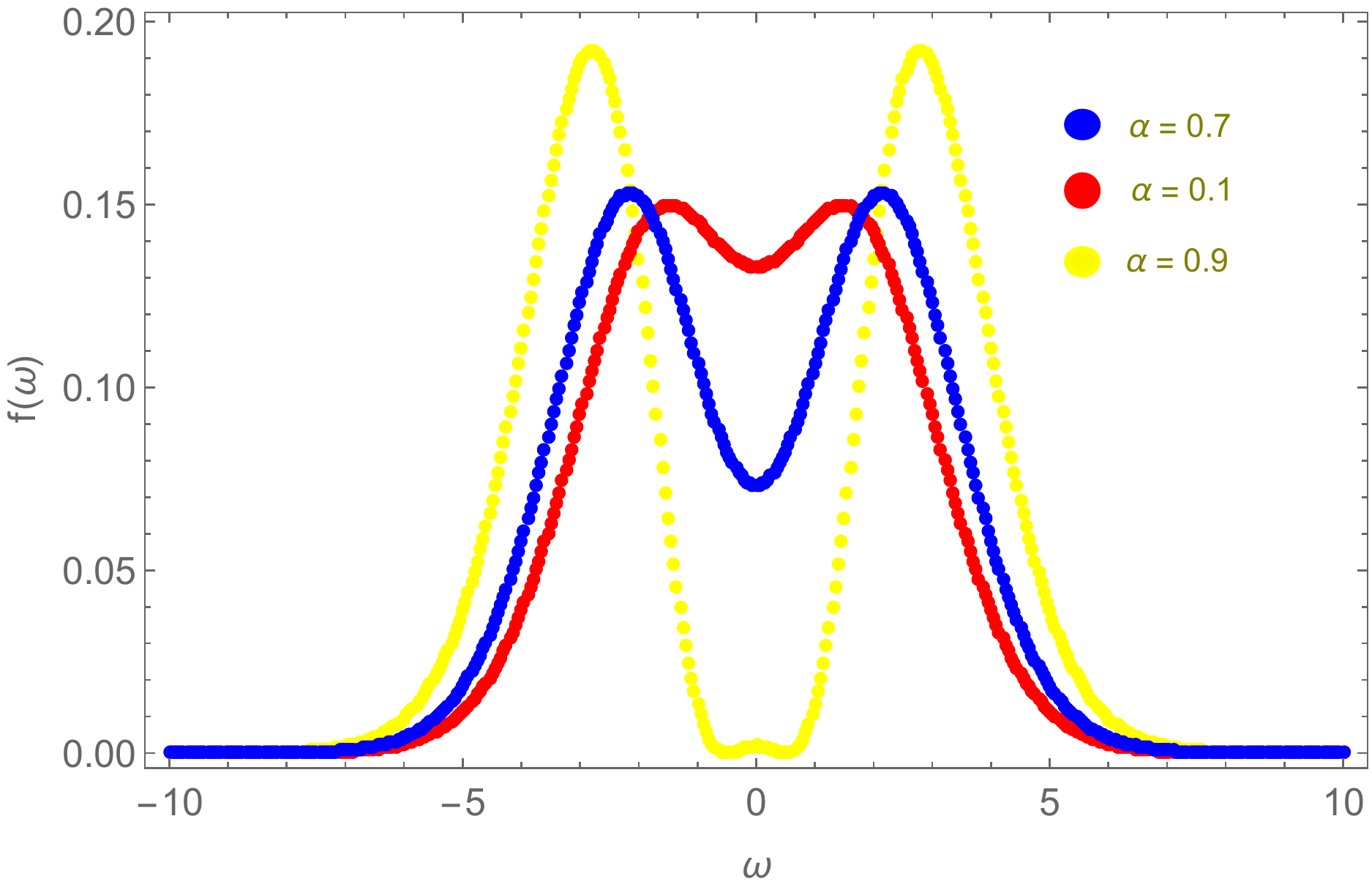}
\end{center}
\caption{Modal fractions for different values of $\protect\alpha $.}
\label{fig1:Fractions}
\end{figure}

The CE for $f(R,T)=R-\alpha \,T^{\,n}$ is plotted in Fig.2 below.

\begin{figure}[h!]
\begin{center}
\includegraphics[width=8.7cm]{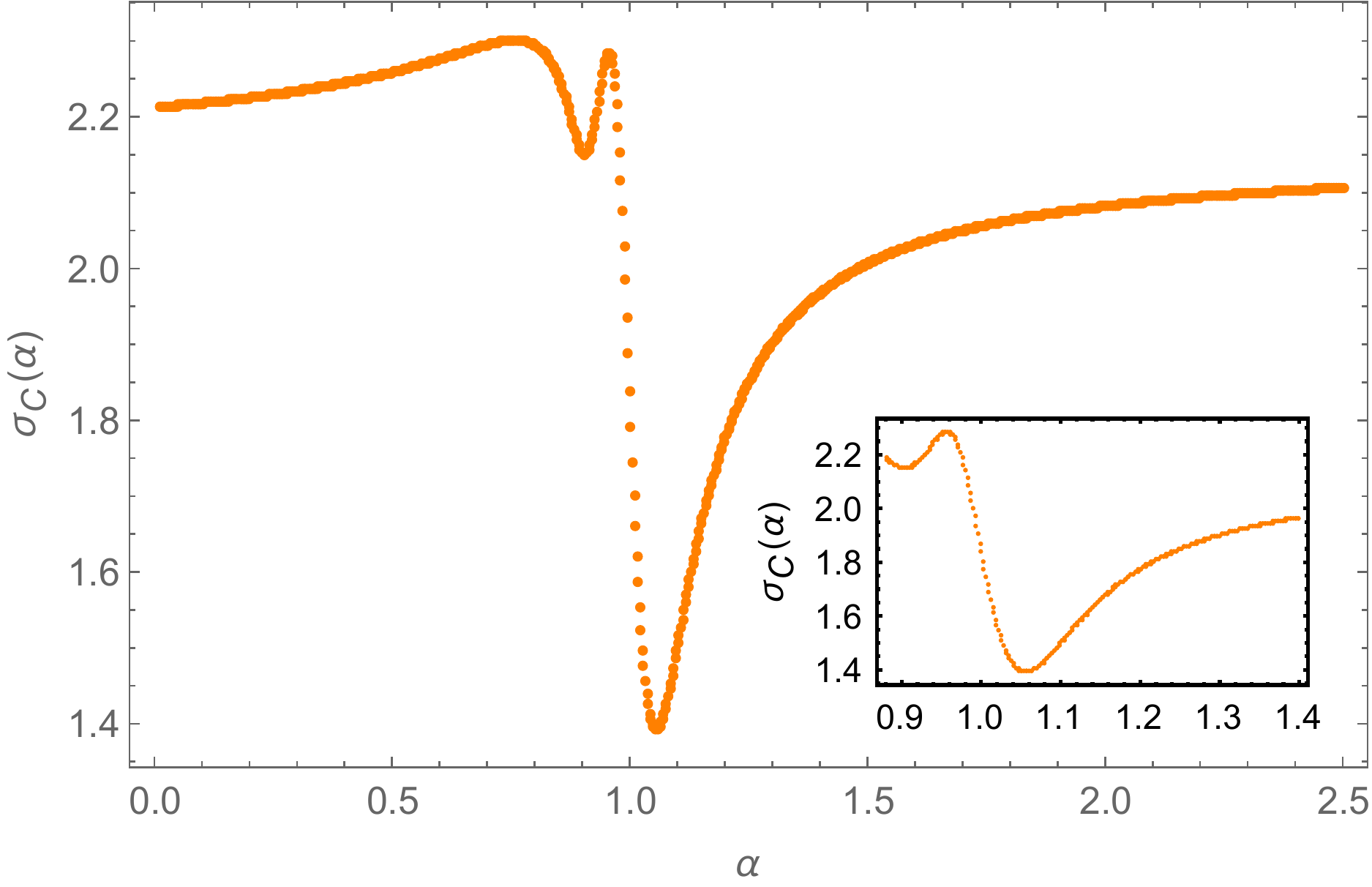}
\end{center}
\caption{Configurational entropy for $f(R,T)=R-\protect\alpha \,T^{\,n}$}
\label{fig2:CE1}
\end{figure}

\subsection{CE for $f(R,T)=R+\protect\beta R^{2}-\protect\alpha T\,$}

Here, the energy density can be put in the form 
\begin{equation}
\rho (y)=\sum\limits_{j=1}^{3}\tilde{A}_{j}\mbox{sech}^{\tilde{\delta}%
_{j}}(ky),  \label{eq.9}
\end{equation}

\noindent where $\tilde{\delta}_{1}=2$, $\tilde{\delta}_{2}=4$ and $\tilde{%
\delta}_{3}=6$, with 
\begin{eqnarray}
\tilde{A}_{1} &\equiv &-\frac{4k^{2}(3+20\beta k^{2})}{4-5\alpha }, \\
\tilde{A}_{2} &\equiv &\frac{4k^{2}(6-5\alpha )(3+232\beta k^{2})}{%
(4-5\alpha )(4-3\alpha )}, \\
\tilde{A}_{3} &\equiv &-\frac{784\beta k^{4}(8-7\alpha )}{(4-5\alpha
)(4-3\alpha )}.
\end{eqnarray}

From the energy density (\ref{eq.9}) we can find the Fourier transform

\begin{equation}
\mathcal{F}(\omega )=\sum\limits_{j=1}^{3}\sum\limits_{\ell =1}^{3}\tilde{A}%
_{j,\ell }\text{ }_{2}\mathcal{G}_{1}[\tilde{\delta}_{j},\tilde{\mu}_{j,\ell
};\tilde{\mu}_{j,\ell }+1;-1],  \label{eq.10}
\end{equation}

\noindent where we are using the following definitions

\begin{eqnarray}
\tilde{\mu}_{j,\ell } &\equiv &[k\tilde{\delta}_{j}-(-1)^{\ell +1}i\omega
]/2k, \\
\tilde{A}_{j,\ell } &\equiv &\frac{2^{\tilde{\delta}_{j}-1/2}\tilde{A}_{j}[k%
\tilde{\delta}_{j}+(-1)^{\ell +1}i\omega ]}{\sqrt{\pi }(k^{2}\tilde{\delta}%
_{j}^{2}+\omega ^{2})}.
\end{eqnarray}

Thus, we can find that 
\begin{equation}
\int_{-\infty }^{\infty }d\omega \left\vert \mathcal{F}(\omega )\right\vert
^{2}=\frac{1}{k}\sum\limits_{j=1}^{3}\sum\limits_{\ell =1}^{3}\frac{2^{%
\tilde{\delta}_{j,\ell }-1}\wp _{j,\ell }\Gamma ^{2}(\delta _{j,\ell }/2)}{%
\Gamma (\delta _{j,\ell })},
\end{equation}

\noindent with 
\begin{equation}
\wp _{j,\ell }\equiv \tilde{A}_{j}\tilde{A}_{\ell }.
\end{equation}

Therefore, we can write

\begin{equation}
f(\omega )=k\left[ \frac{\sum\limits_{j,\ell ,m,n=1}^{3}\tilde{A}_{j,\ell }%
\tilde{A}_{m,n}^{\ast }\text{ }_{2}\mathcal{G}_{1}^{(j,\ell )}\text{ }_{2}%
\mathcal{G}_{1}^{(m,n)\ast }}{\sum\limits_{j,\ell =1}^{3}\frac{2^{\tilde{%
\delta}_{j,\ell }-1}\wp _{j,\ell }\Gamma ^{2}(\tilde{\delta}_{j,\ell }/2)}{%
\Gamma (\tilde{\delta}_{j,\ell })}}\right] ,
\end{equation}

\noindent where 
\begin{eqnarray}
_{2}\mathcal{G}_{1}^{(j,\ell )} &=&_{2}\mathcal{G}_{1}[\tilde{\delta}_{j}, 
\tilde{\mu}_{j,\ell };\tilde{\mu}_{j,\ell }+1;-1], \\
_{2}\mathcal{G}_{1}^{(m,n)} &=&_{2}\mathcal{G}_{1}[\tilde{\delta}_{m},\tilde{
\mu}_{m,n};\tilde{\mu}_{m,n}+1;-1].
\end{eqnarray}

Figures 3-4 below show the modal fraction for $\alpha =0.1$ and $\alpha =0.98
$, respectively. From that figures, we can note that the $\alpha $
parameters induce the formation of two peaks which are symmetric in relation
to the origin. 
\begin{figure}[h]
\begin{center}
\includegraphics[width=8.7cm]{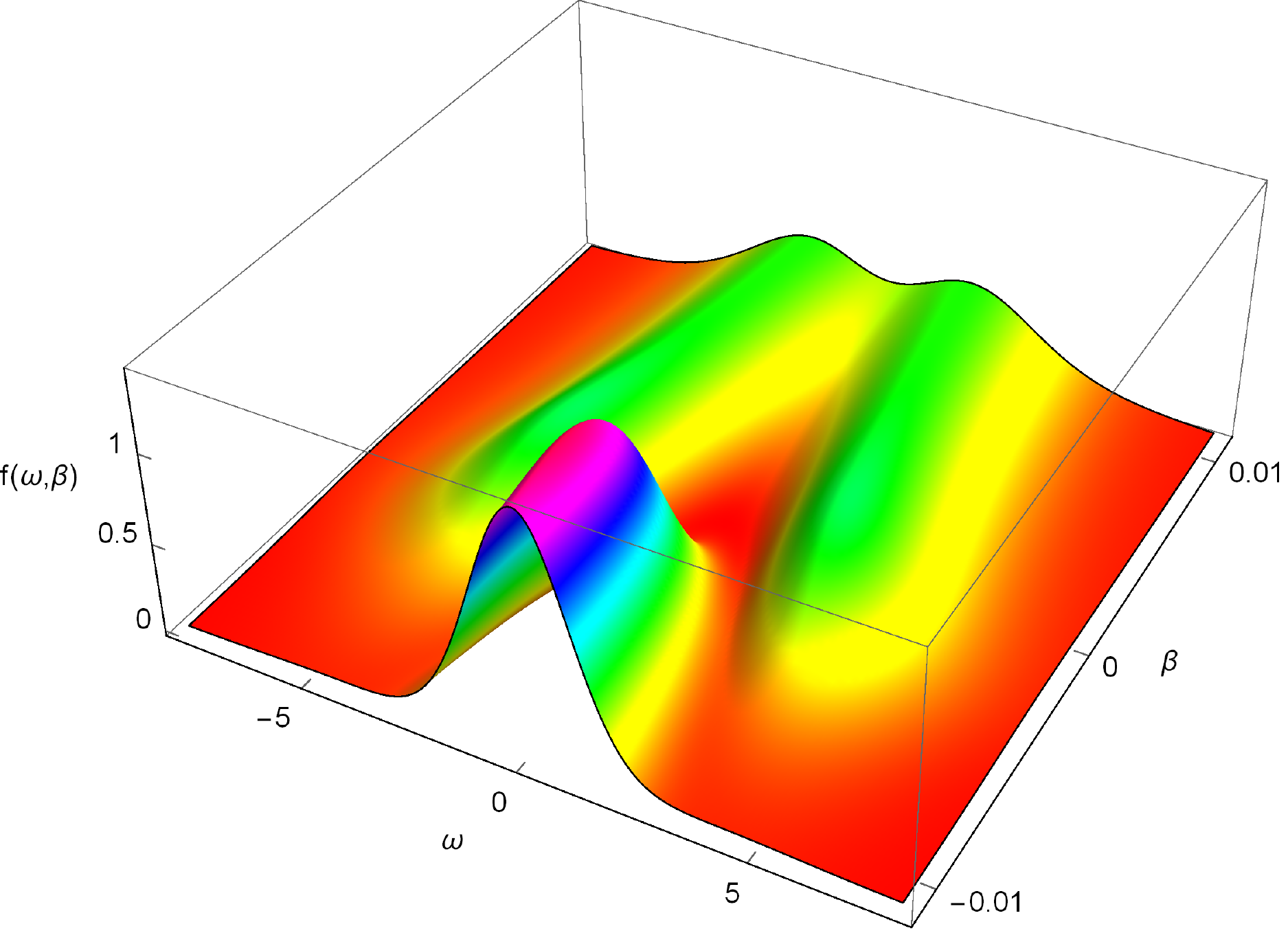}
\end{center}
\caption{Modal fraction with $\protect\alpha =0.1$.}
\label{fig3:MF2a}
\end{figure}

\begin{figure}[h]
\begin{center}
\includegraphics[width=8.7cm]{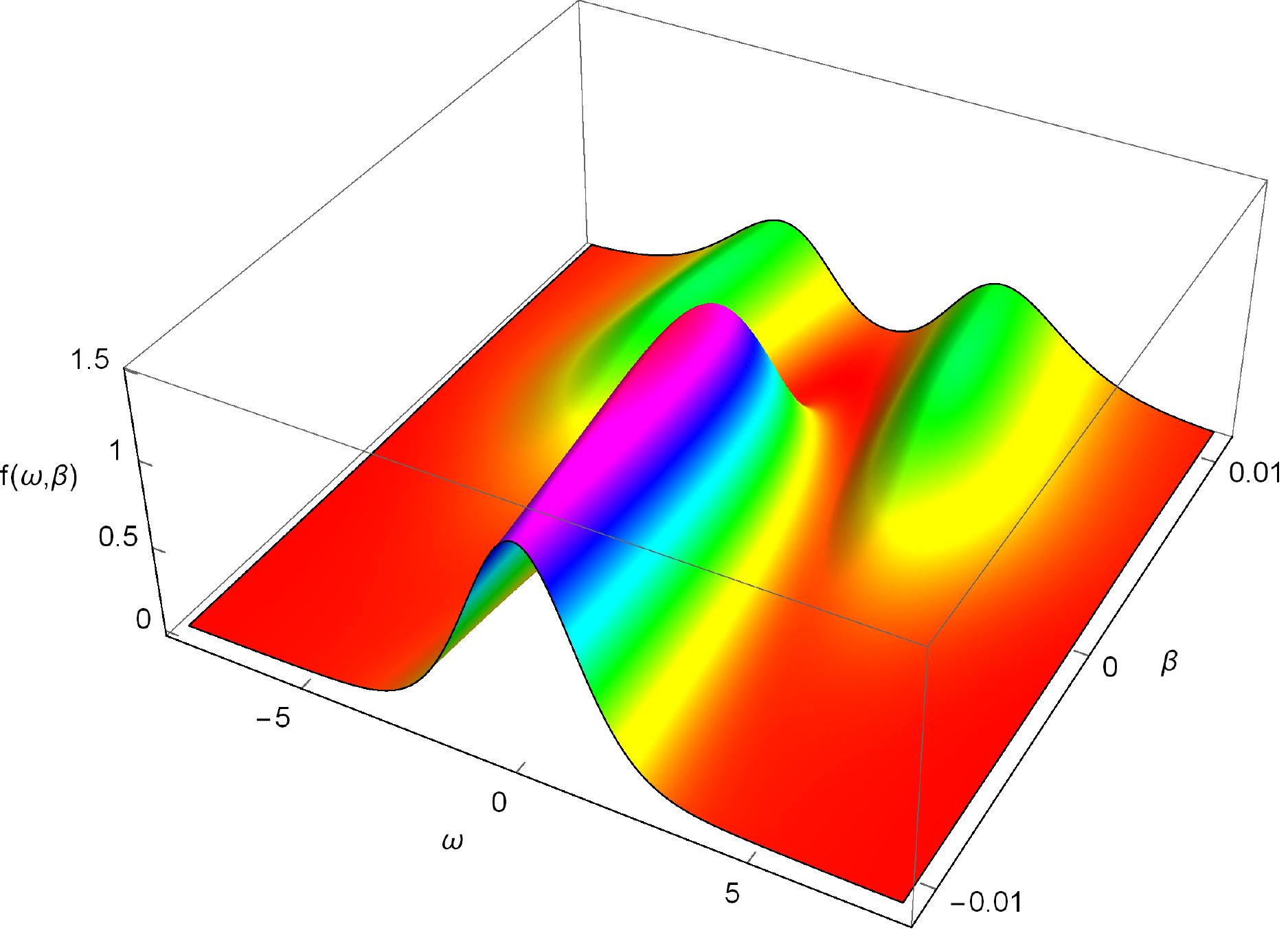}
\end{center}
\caption{Modal fraction with $\protect\alpha=0.98$.}
\label{fig4:MF2b}
\end{figure}

In Fig.5 below we plot the CE for $f(R,T)=R+\beta R^{2}-\alpha T\,$.

\begin{figure}[h]
\begin{center}
\includegraphics[width=8.7cm]{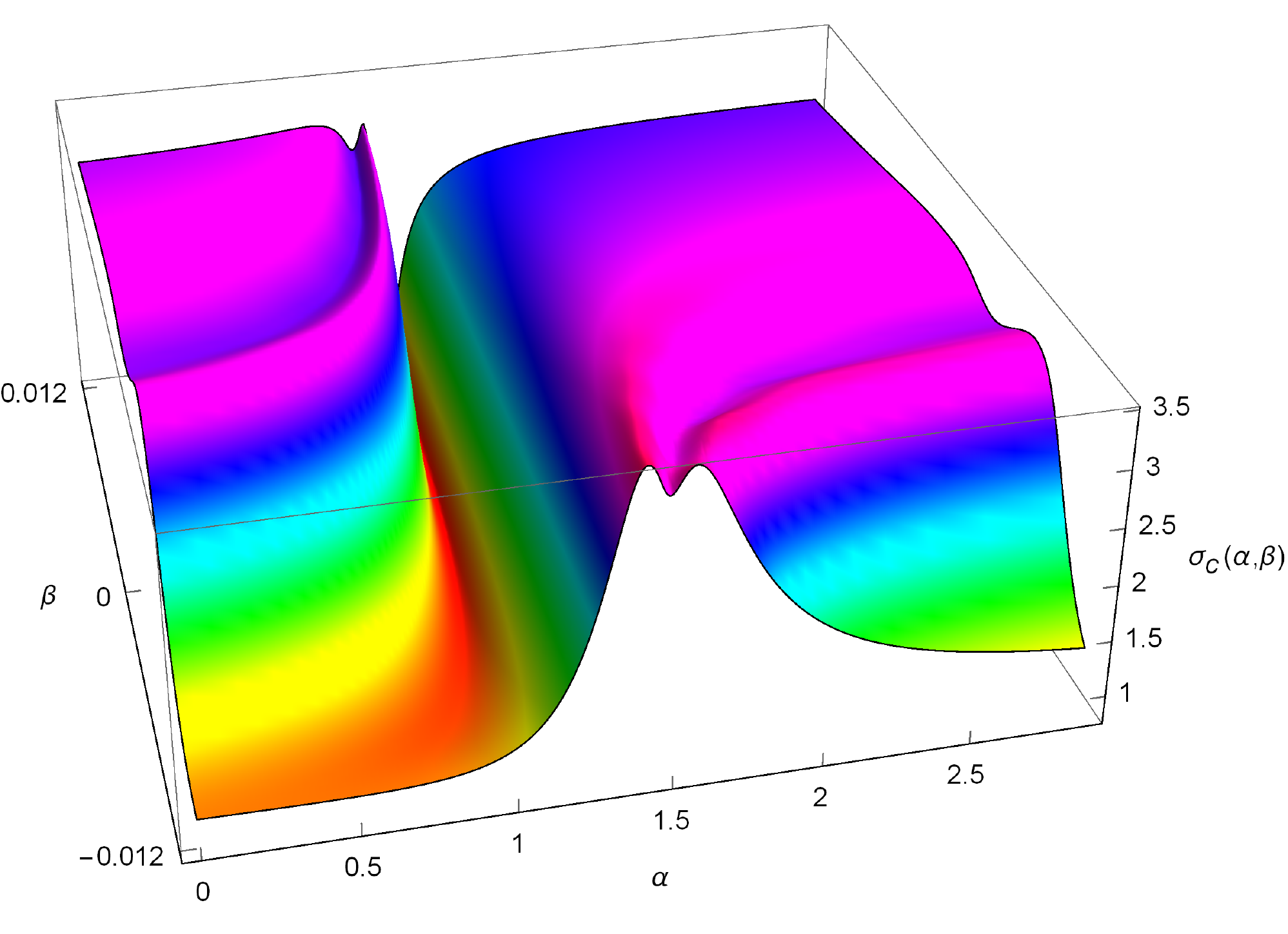}
\end{center}
\caption{Configurational entropy for $f(R,T)=R+\protect\beta R^{2}-\protect%
\alpha T\,$}
\label{fig5:CE2}
\end{figure}

\section{Conclusions}

In this work we investigated $f(R,T)$ braneworld models in the CE context.
We showed, by means of this information-theoretical measure, that a stricter
bound on the parameter of $f(R,T)$ theories of gravity arises from the CE.
We found that these bounds are characterized by a valley region in the CE
profile, where the entropy is minimal.

From inspection of such results, we can see that in the case where $%
f(R,T)=R-\alpha \,T^{\,n}$, the CE shows a rich structure for varying $%
\alpha $. In this case, there is a sharp minimum at the value $\alpha \simeq
1.07$, where the corresponding CE is given by $\sigma _{c}\simeq 1.4$. This
result leads us to conclude that the CE can be used in order to extract a
rich information about the structure of the configurations, which is clearly
related to their profiles. Therefore, we found that the best ordering for
the solutions are those given by $\alpha \simeq 1.07$.


Moreover, when $f(R,T)=R+\beta R^{2}-\alpha T\,$, the minimal CE corresponds
to a valley region where the best values for $\alpha $ and $\beta $ are
localized in the bottom of the valley. In this way, the
information-theoretical measure of generalized theories of gravity, such as $%
f(R,T)$ theories, opens a new window to probe situations where the
parameters responsible for the control of the theory are, at first sight,
arbitrary. In this case, the CE selects the best values.

Thus, in this work, we can conclude that the CE provides a complementary
perspective to investigate generalized theories of gravity such as
Gauss-Bonnet, Weyl and Brans-Dicke theories, as well as several other
modified theories of gravitation. Another interesting line of investigation
in which CE can play an important rule is the evolution of the domain walls
in Euclidean space \cite{Rafael-Pedro-Roldao}. In this case, the
information-theoretical measure can be used as a discriminant to understand
the phase transition processes.

\acknowledgements

RACC thanks to UFABC and CAPES for financial support, and PHRSM thanks to
FAPESP for financial support.

\end{document}